\documentstyle[preprint,aps,epsf]{revtex}

\newcommand{\etal}{{\em et al. }}                
\newcommand{\beq}{\begin{equation}}
\newcommand{\eeq}{\end{equation}}
\newcommand{\beqar}{\begin{eqnarray}}
\newcommand{\eeqar}{\end{eqnarray}}
\newcommand{\beqars}[1]{\begin{eqnarray*}{#1}}
\newcommand{\eeqars}{\end{eqnarray*}}
\newcommand{\plb}[1]{Phys.~Lett.~B {#1}}

\newcommand{\pdd}{P.~Danielewicz}

\def\){\right) }
\def\({\left( }
\def\]{\right] }
\def\[{\left[ }

\def\rj{$\lbrace r_j \rbrace_{i =1 , \ldots , N}\,$}

\def\lapproxeq{$\,\,$\raisebox{-.5ex}{$\stackrel{<}{\scriptstyle\sim}$$\,\,$}}

\begin{document}
\draft

\title{
\begin{flushright}{\rm MSUCL-1086}\\[9mm]\end{flushright}
Optimized Discretization of Sources\\
Imaged in Heavy-Ion Reactions
}

\author{
David A.\ Brown\thanks{e-mail:
dbrown@nscl.msu.edu}
and
Pawe\l~Danielewicz\thanks{e-mail:
danielewicz@nscl.msu.edu}
}
\address{
National Superconducting Cyclotron Laboratory and\\
Department of Physics and Astronomy, Michigan State University,
\\
East Lansing, Michigan 48824, USA\\
}
\date{\today}
\maketitle

\begin{abstract}
We develop the new method of optimized discretization for imaging
the relative source from two particle correlation functions.
In this method, the~source resolution depends on the relative
particle separation and is adjusted to available data and
their errors.  We test the method by restoring
assumed pp sources and then apply the method to pp and IMF data.
In reactions below 100~MeV/nucleon, significant
portions of the sources extend to large distances ($r > 20$~fm).
The~results from the imaging
show the inadequacy of common Gaussian source-parametrizations.
We establish a simple relation between the height of the pp
correlation function and the source value at short distances,
and between the height and
the proton freeze-out phase-space density.
\end{abstract}


\narrowtext

\section{Introduction}
Imaging techniques are used in many diverse
areas such as geophysics, astronomy, medical diagnostics, and police
work.  The goal of imaging varies widely from determining the density
of the Earth's interior to reading license plates from
blurred photographs to issue speeding fines.  A typical linear imaging
problem involves extracting a source from measured data, where the data
is represented as the convolution of a kernel with a source.  An example
of this is imaging the relative source function $S_{{\bf P}}({\bf r})$ from
the two--particle correlation function, $C_{{\bf P}}({\bf q})$,
often measured in nuclear reactions~\cite{bro97}:
\beq
C_{{\bf P}}({\bf q}) -1 \equiv {\cal R}_{{\bf P}}({\bf q}) =
\int d{\bf r} \,
K({\bf q},{\bf r}) \, S_{{\bf P}} ({\bf r}) \, .
\label{K}
\eeq
Here ${\bf P}$ is the total momentum of the particle pair and
the kernel $K$ is is equal to~\cite{koo77,pra90,dan92}
\beq
K({\bf q},{\bf r}) = |\Phi_{{\bf q}}^{(-)}({\bf r})|^2-1  \, .
\label{KPhi}
\eeq
The wavefunction $\Phi^{(-)}$ describes the propagation of the pair from a cm
separation ${\bf r}$
out to the detector at
infinity,
where
relative momentum ${\bf q}$ is reached.  In the semiclassical limit,
the~relative source $S_{\bf P}$ for any two particles
represents the
probability distribution for emitting the the two particles with a
separation ${\bf r}$, in their center of mass.

On the face of it, imaging appears straightforward:  in the case of the
two--particle correlation function, one could just discretize $C_{{\bf P}}$
and $S_{{\bf P}}$ and invert the resulting matrix equation.  However, in
practice this may not work as small variations in data, within
statistical or systematic errors, can generate huge changes in
the
imaged source.   This stability problem is well known in other fields
and vast literature exists on its
resolution~\cite{boe85,smi85,pre92,bal80,bac68}.
In fact, many imaging  problems~\cite{ber80,pre92} that are
now routinely solved would be considered
ill-posed in the sense of Hadamard, who discussed stability
in inversion as early as in~1923~\cite{had23}.
A~key element in stabilizing the inversion process was found by
Tikhonov~\cite{tik63} who showed that inversion problems can be
regularized by using known constraints satisfied by the~source.
Thus,  in~optical imaging (e.g.\ in astronomy)
the~most common method is the
maximum entropy method~\cite{ski85,bev85,pre92}.
The~maximum entropy method works by
using the~most likely (i.e.\ uniform) image at scales below
the resolution limit to~stabilize the restored image  at larger
scales. In this method, the image is constrained to be positive.
In this paper, we describe a method for imaging the
relative source function from the two--particle correlation
function, without explicit constraints.  Our new method is
similar in spirit to the maximum entropy method in that we optimize
the scale of the kernel~$K$ and then use the optimized scale to
restore the actual source (i.e.\ the entire image).

In~a~recent paper~\cite{bro97}, we demonstrated that
it is possible to image the relative sources in heavy-ion
reactions.  This represented a step forward
as, in~the past, the~integral relation between the correlation functions and
the sources was only used to determine the gross
features of the sources, such as representative radii.
In~\cite{bro97}
we~put forward two methods of imaging sources from correlations.
In~the~first of these methods, we used the fact that, when
mutual interactions could be ignored or easily corrected
such as with photons or pions, the~kernel is dominated by
Bose--Einstein
interference.  The~kernel then becomes a~cosine and imaging
turns to inverting  a Fourier cosine transform.
The~method allowed us to determine source values for any separation
within a~certain range.  In~the~second method, we simply discretized
the source at {\em equal} intervals. This method is more general and can be
applied to such cases as proton-proton~(pp) correlations.
We did not discuss the~stability of either of these methods.

The~question that we now ask is whether discretizing the source
at equal intervals is optimal in our second, more general,
imaging method~\cite{bro97}.  For example, in the pp case,
the correlation function is dominated by the Coulomb interaction
at low-relative momenta and by the strong interaction and
antisymmetrization at intermediate momenta.  The~different
momentum regions should give access to large and short distances
within the source, respectively, with the resolution decreasing
at the large distances, rather than being fixed.
This suggests that, by varying the size of the discretization
interval, we can optimize the kernel to best restore a given source.
Since we do not know the source ahead of time, we must specify an
``model source'' in order to choose the best kernel.  This kernel
is then used in the actual inversion process.

We illustrate the seriousness of the stability problems by
constructing a test correlation function, with errors, out
of an assumed source and attempting to restore the source.
Next, we demonstrate how our method stabilizes the inversion.
Following this, we apply the new method to the pp
data~\cite{gon90,gon91} already
analyzed in~\cite{bro97}.  We assess whether the sources are
positive definite, which would allow for their semiclassical
interpretation.
We compare the results to results from the transport model of
reactions~\cite{dan95}.
The~pp~sources exhibit significant variation with total
pair momentum gate.
We investigate the integrals of the source functions
over imaged regions and
we show how delicate the analysis of
NN correlations can be in the different gates.
Finally, we apply our method to analyze
intermediate-mass-fragment (IMF) correlations~\cite{ham96}.
The~IMF~sources exhibit significant variation with
bombarding energy.

\section{Imaging}

In this section we discuss the general imaging technique.
We begin with equations~(\ref{K}) and~(\ref{KPhi}).   Since our
goal
is to restore the source function $S_{{\bf P}}$, we should discuss
briefly what it is.  The~source $S_{{\bf P}}$ is proportional to
a~Wigner transform
of a~quantal expectation value~\cite{dan92,ber94}, and  in~the
semiclassical limit, the~source represents the distribution
of relative cm separation of the last collisions of the two
particles.  In this limit, it may be expressed in terms of
production and absorption rates~\cite{dan92,bro97} and, by nature,
it is positive definite.  Generally, the~source is
normalized to~1.  When pair emission is significantly distorted
in the vicinity of the source, e.g.\ due
to the Coulomb interaction with the source, those distortions
may be accounted for either in the wavefunction in the kernel or
absorbed into the definition of the source.  If~the latter is
the case, or distortions are simply negligible,
the~functions in~(\ref{K})
may be expanded into spherical harmonics and relations between
angular coefficients in the expansion follow~\cite{bro97}.
In particular, the~angle-averaged
correlation function is related to the
angle-averaged source through the~angle-averaged kernel:
\beq
C_{{\bf P}}( q ) -1 \equiv {\cal R}_{{\bf P}}( q ) =
4 \pi \int_0^\infty d{r} \, r^2 \,
K_0 ( q , r) \, S_{{\bf P}} ( r ) \, .
\label{Kav}
\eeq

The form of the kernel depends on the particle pair.
In the pion-pion case the kernel is $K({\bf q},
{\bf r}) = \cos{(2 {\bf q} \cdot {\bf r})}$, given that the
mutual Coulomb interaction is corrected for in~$C_{{\bf P}}$.
The~angle-averaged kernel
follows simply as $K_0 ( q, r) = \sin{(2 q r)}/(2qr)$.
For these kernels,
the~source~$S_{{\bf P}}$ is an inverse Fourier
transform of~$C$~\cite{bro97}.
For the angle-averaged source one finds:
\beq
r \, S_{{\bf P}}(r) = {2 \over \pi^2} \, \int_0^{\infty} dq \,
q \, \sin{(2qr)} \, (C_{{\bf P}} (q) - 1) \, .
\label{rSr}
\eeq
In practice, the~integration in (\ref{rSr}) is cut off at
some suitably chosen value of $q_{\rm max}$.  This limits, from
above, the region where the correlation in~(\ref{Kav}) is
dominated by two-pion interference.
In~the proton-proton~(pp) case, the~angle and spin averaged
kernel is~\cite{bro97}
\beq
K_0 (q, r) = {1 \over 2} \sum_{j s \ell \ell'} (2 j +1) \,
\left( g_{js}^{\ell \ell'} (r) \right)^2 -1 \, ,
\eeq
where $g_{js}^{\ell \ell'}$ is the radial wave function
with outgoing asymptotic
angular momentum~$\ell$.
In~the classical limit of correlations of purely Coulomb
origin, such as investigated between IMFs~\cite{kim91},
the~kernel is
\beq
K_0 (q,r) = \theta(r - r_c) \,
\left( 1 - r_c/r \right)^{1/2} - 1 \, ,
\label{KCou}
\eeq
with the distance of
closest approach $r_c (q) = 2 \mu \, Z_1 \, Z_2 \, e^2/q^2$.

Typically in an~experiment, the correlation function~$C_{{\bf P}}$ is
determined at discrete values of the magnitude of relative
momentum $\lbrace q_i \rbrace_{i = 1, \ldots, M}$ for
directionally-averaged function, or on a~mesh
in the momentum space $\lbrace {\bf q}_i \rbrace_{i = 1,
\ldots, M}$  when no averaging is done.  With each
determined value~$C_i^{\rm exp}$ some
error $\sigma_i$ is
associated.  It is this set of
values $\lbrace C_i^{\rm exp} \rbrace_{i=1, \ldots, M}$, that
we use in determining the source function. In the 3-dimensional
case we may introduce a~rectangular mesh in the space of
relative particle separation and assume that the source
function is approximately constant within different cells of
the mesh.  In the angular expansion of~\cite{bro97},
we may also assume
that the spherical expansion coefficients vary slowly within the
$r$-intervals.
In~the angle-averaged case, discretization amounts simply
to the representation $S
\simeq \sum_{j=1}^N S_j \, g_j(r)$, where~$N$ is the number of
intervals in~$r$, $g_j(r) = 1$ for $r_{j-1} < r <
r_j$, and $g_j(r) = 0$,  with $r_j = j \,
\Delta r$.  On~inserting
a~discretized form of~$S$ into Eq.~(\ref{K}) or~(\ref{Kav}),
we find a~set of equations for
the correlation functions $\lbrace C_i^{\rm th}
\rbrace_{i=1, \ldots, M}$, in~terms of $\lbrace S_j
\rbrace_{j=1, \ldots, N}$, \beq
C_i^{\rm th} - 1 \equiv {\cal R}_i^{\rm th} = \sum_{j=1}^N
K_{ij} \, S_j \, ,
\eeq
where, in the angle-averaged case,
\beq
K_{ij} = 4 \pi
\int_{r_{j-1}}^{r_j} dr \, r^2 \, K(q_i, r) \, .
\label{Kij}
\eeq
The~source
values may be searched for by a~$\chi^2$ minimization
\beq
\sum_{i=1}^M {(C_i^{\rm th} - C_i^{\rm exp})^2 \over \sigma_i^2
} = {\rm min} \, .
\label{chi}
\eeq

If we do not constrain the space within which we search
for~$\lbrace S_j \rbrace_{j=1, \ldots, N}$, then a~set of
linear equations for the values follows by
differentiating~(\ref{chi}) with respect to~$\lbrace S_j
\rbrace_{j=1, \ldots, N}$,
\beq
\sum_{ij} {1 \over \sigma_i^2} \left(K_{ij} \, S_j -
{\cal R}_i^{\rm exp} \right) K_{ik} = 0 \, ,
\eeq
or in a matrix form
\beq
K^\top \, B \, (K \, S - {\cal R}^{\rm exp}) = 0 \, ,
\eeq
where $B_{il} = \delta_{il}/\sigma_i^2$.
This matrix equation can be solved for $S$:
\beq
S = \left( K^\top \, B \, K \right)^{-1} \, K^\top \, B \,
{\cal R}^{\rm exp} \, .
\label{S=}
\eeq

\section{Image Stability}
\label{stability}

Our discussion of the stability of the imaged source function begins
with the errors of $\lbrace S_j\rbrace_{j=1, \ldots, N}$
that we determine from (\ref{S=}),
\beq
\Delta^2 S_j = \left( K^\top \, B \, K \right)^{-1}_{jj} \, .
\label{DS}
\eeq
The $N \times N$ matrix $K^\top \, B \, K$ in (\ref{S=}) is
symmetric, positive definite and may be diagonalized,
\beq
\left( K^\top \, B \, K \right)_{kj} \equiv \sum_{i=1}^M
{1 \over \sigma_i^2} \, K_{ik} \, K_{ij}
= \sum_{\alpha = 1}^N \lambda_\alpha \, u_i^\alpha \,
u_j^\alpha \, ,
\label{diag}
\eeq
where $\lbrace u^\alpha \rbrace_{\alpha = 1, \ldots, N}$ are
orthonormal and
$\lambda_\alpha \ge 0$.  With (\ref{DS}) and~(\ref{diag}),
the~square errors for individual values of~$S$ are
\beq
\Delta^2 S_j = \sum_\alpha {(u_j^\alpha)^2  \over
\lambda_\alpha} \, .
\label{DSl}
\eeq

We see in (\ref{DSl}) that the errors of the source
diverge (or the~inversion problem becomes unstable) if~one or
more of the eigenvalues~$\lambda$ approaches zero.  In particular,
this happens when~$K$ maps an~investigated spatial region to
zero. A~specific case is when one of of the particles is neutral
so $|\Phi|^2\approx 1$,
cf.\ Eq.~(\ref{KPhi}).  Moreover, instability can arise when one
demands too high a resolution for a given set of measurements.
In such a~case, what might happen is that~$K$ smoothes out
variations
in $S$, so we lose this information in the correlation function.
If we then try to restore $S$, we find that
we cannot restore~$S$ uniquely at high resolution. However at lower
resolution, we might still be able to restore it.
Unlike typical numerical methods,
we need a singular rather than a smooth
kernel~\cite{pre92,deh80} for our inversion problem to be tractable.
Finally, a~$\lambda$ close to zero can
be reached by accident for an~unfortunate choice of
$\lbrace
r_k \rbrace_{k = 1, \ldots, M}$ for a  given  measurement.

\begin{figure}
\vspace*{.4in}
\begin{center}
\epsfxsize=5.5in 
\epsffile{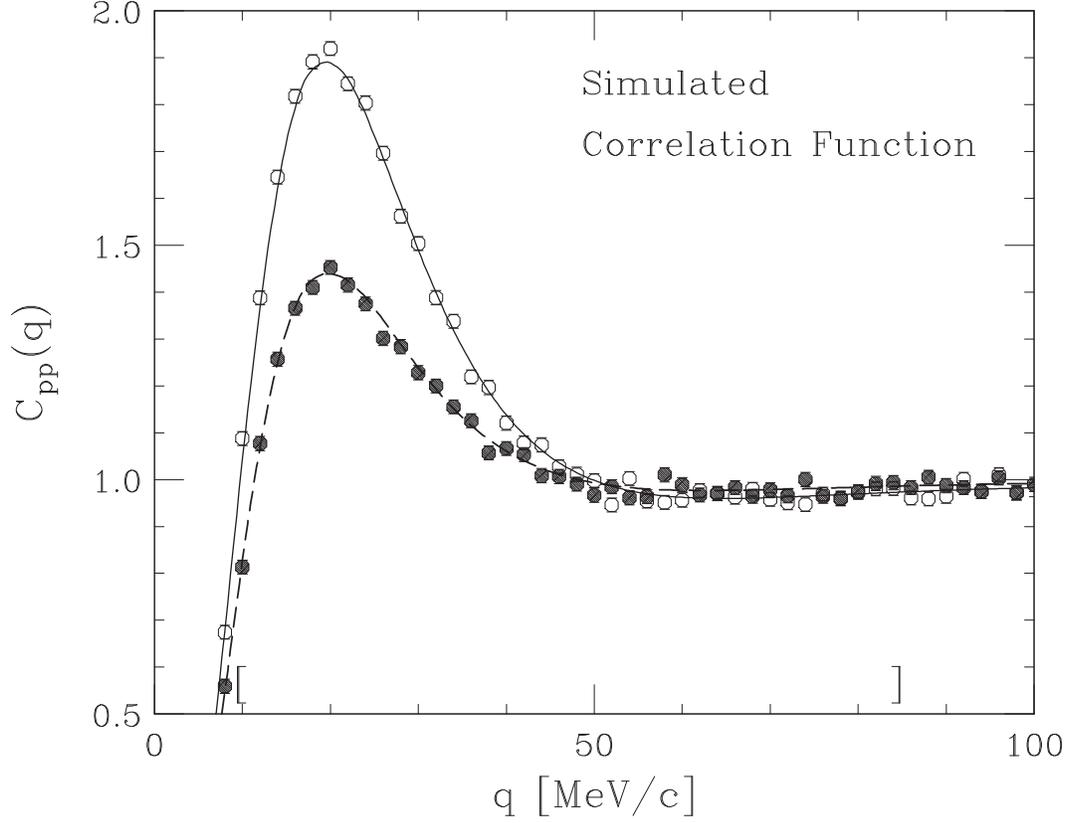}
\vspace*{.4in}
\end{center}
\caption{The solid line is the correlation function from the source
in Eq.~(\protect\ref{SG}) and the dashed line is from the source in
Eq.~(\protect\ref{SGo}).
We obtained the wavefunctions in the kernel in~(\protect\ref{Kav})
by solving the Schr\"odinger equation with the REID93
potential~\protect\cite{sto94}. The symbols
are the values of the correlation functions with added
random errors of rms magnitude~0.015.  The~square brackets
above the horizontal axis indicate the range of~$q$ we used to restore
source.}
\label{simcor}
\end{figure}

\begin{figure}
\vspace*{.2in}
\begin{center}
\epsfxsize=5.5in 
\epsffile{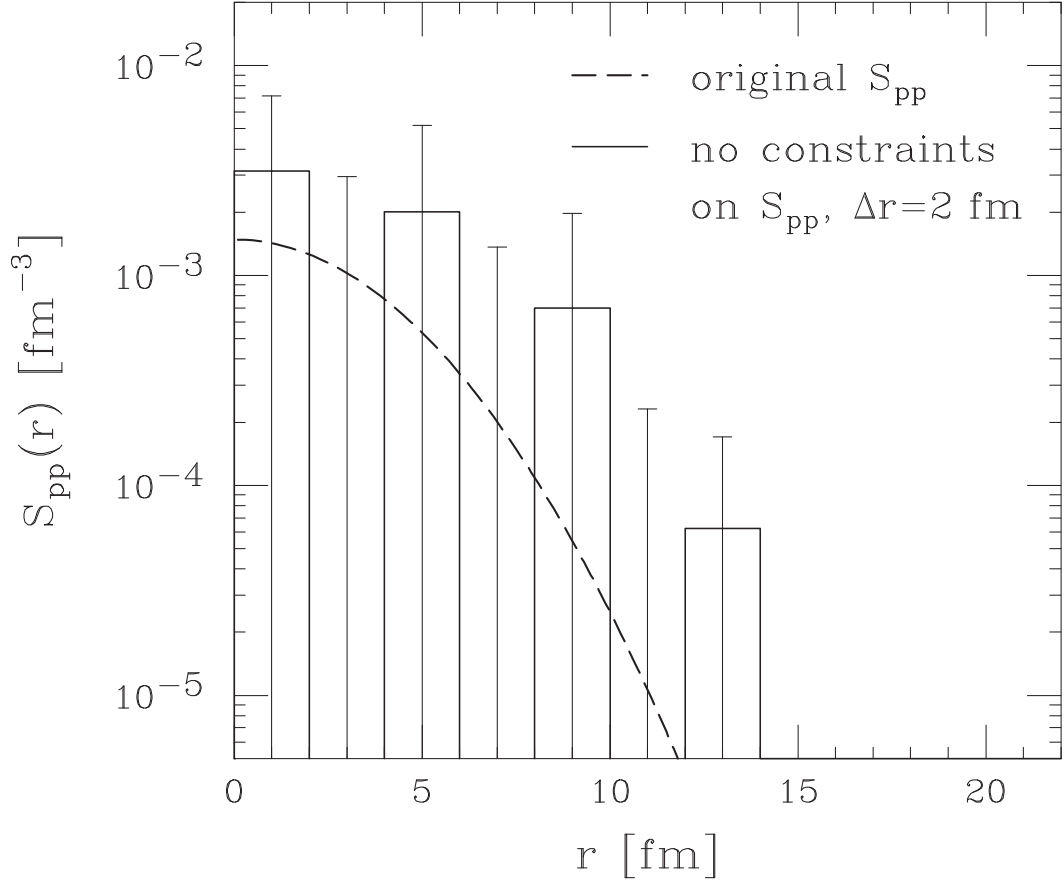}
\vspace*{.4in}
\end{center}
\caption{The solid histogram is the relative pp source
function~$S$ restored from the simulated correlation function
in Fig.~\protect\ref{simcor} indicated there by the open
symbols.
The dashed line is the original source function in~(\protect\ref{SG})
that we used to generate the correlation function.
We employed fixed-size intervals of $\Delta r = 2$~fm
for discretizing the source function and we imposed
no constraints on~$S$.}
\label{nocon2}
\end{figure}

To illustrate how serious the issue of stability can be, we
take a relative pp source of a~Gaussian form
\beq
S(r) = {1 \over (2 \pi R_0^2)^{3/2}} \exp{\left(- \,
{r^2 \over 2 R_0^2} \right)} \, ,
\label{SG}
\eeq
and
generate a~correlation function~$C$ at relative momenta $q$
separated by $\Delta q = 2$~MeV/c.  We
use the folding~(\ref{Kav})
with the wavefunctions in the kernel calculated by solving the
Schr\"odinger equation with the regularized Reid soft-core
potential REID93~\cite{sto94}.  This simulated source is shown in
Fig.~\ref{simcor}.  We~take $R_0 = 3.5$~fm
in the source and we add random Gaussian-distributed errors
to the correlation function from the folding.  The~rms
magnitude of the error is~0.015, which is representative of the
pp data of
Ref.~\cite{gon90} analysed in~\cite{bro97}.  We then attempt to
restore the source by discretizing it with~7 intervals of
fixed size, $\Delta r = 2$~fm, for $r
= (0-14)$~fm.  We use a $q$--interval similar to the one used in
Ref. \cite{bro97}, i.e. $10$~MeV/c~$<q<86$~MeV/c.
(Note that a naive Fourier-transform
argument suggests that a~number larger than~7 of
$r$-intervals narrower than 2~fm could be used in the
inversion.)  Results from the restoration according
to Eqs.~(\ref{S=})
and~(\ref{DS}) are shown in Fig.~\ref{nocon2}, together with
the original source function~(\ref{SG}).  Clearly the
errors for restored source far exceed the original source
function.
In~fact, every second value of the restored source is
negative.  Incidentally, this is
one of our more fortunate simulations, since all of the errors
actually fit on the logarithmic plot.

\begin{figure}
\vspace*{.2in}
\begin{center}
\epsfxsize=5.5in 
\epsffile{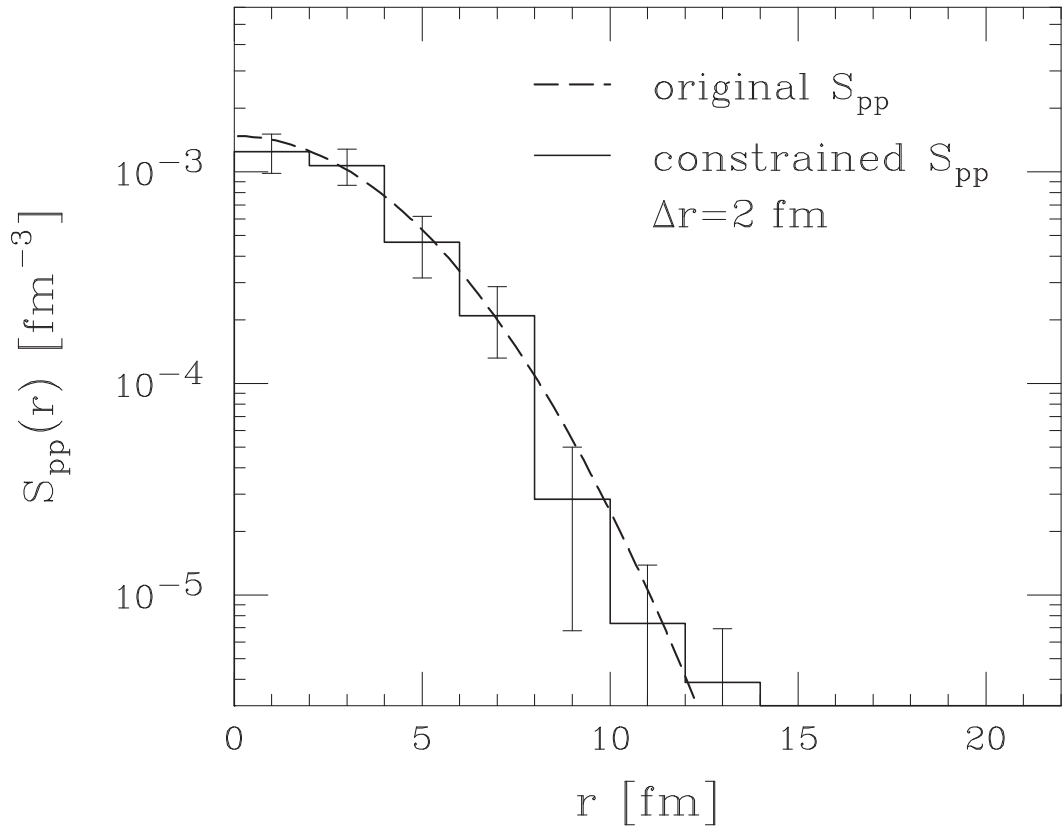}
\vspace*{.4in}
\end{center}
\caption{
This plot is the same as Fig.~\ref{nocon2}, except that the
restored
source is constrained to be positive and its integral is
restricted to be less than one.
}
\label{con2}
\end{figure}

We next illustrate the dramatic stabilizing effect that
the constraints have on the imaging,
as it was first discussed by Tikhonov~\cite{tik63}.
We carry out the inversion using the same
correlation function and errors that we used
for Fig.~\ref{nocon2} (see Fig.~\ref{simcor}).
We impose the constraints that the imaged source is positive
definite, i.e. $S_j \ge 0$, as  expected in the
semiclassical limit, and that the source is
normalized: within the restored region
$1 \ge 4 \pi \int_0^{r_N} dr \, r^2 \, S \approx 4 \pi
\sum_{j=1}^N S_j \int_{r_{j-1}}^{r_j} dr \, r^2$.
Following the general strategies~\cite{dag95} for
estimating values with errors and under constraints,
we carry out our imaging by sampling the values of the
correlation
of function according to the errors $\lbrace \sigma_i
\rbrace_{i = 1, \ldots, N}$ (equal to 0.015 in our case) and
by applying~(\ref{S=}).  This amounts to the
replacements in~(\ref{S=}): $C_i
\rightarrow C_i + \sigma_i \, \xi_i$, $i = 1,
\ldots, N$, with $\xi$'s drawn from the standard normal
distribution.  We~accept only those samplings where the
constraints
are met.  With these samplings, we calculate the average
source values
and the average dispersions.  The~results are shown in
Fig.~\ref{con2} together with the original source. They
now compare favorably to the original source.

In~the maximum entropy
method~\cite{pre92}, as we stated, the~constraint of positive
source definiteness is implicit.  Another constraint
one might use~\cite{tik63,ber80,deh80} is an assumption of
smoothness of the source, permitting inversion with
more points in the image then there are in the data.
Cutting off the integral in the Fourier-transform
method~\cite{bro97} at $q_{\rm max}$ implies the constraining
assumption
that~$S$ varies slowly on the scale of $1/(2q_{\rm max})
\approx 1.2$~fm in~\cite{bro97}, which is reasonable given the
range of strong interactions.
Finally we comment that, in imaging terms, the common
Gaussian parametrization of
sources in heavy-ion collisions is a very extreme constraint
for stabilizing the inversion.

\section{Optimized Discretization}

While imposing constraints on the source may stabilize
the inversion, we have developed an~imaging method that can yield
very satisfactory
results even without any constraints.
Indeed, one may want to see directly whether the
data are consistent with positive definite sources.
The~essential step is to vary the size of the $r$--intervals to
minimize relative error of the source.

Thus, the~first stage of analysis involves the values of
relative momenta
$\lbrace q_i \rbrace_{i =1 , \ldots , M}$, where
correlation function was measured, and errors on
these measurements
$\lbrace \sigma_i \rbrace_{i =1 , \ldots , M}$, but not the
values themselves $\lbrace C_i \rbrace_{i =1 , \ldots , M}$.
Specifically, we~vary the edges of the intervals
for source discretization,
$\lbrace r_j \rbrace_{i =1 ,
\ldots , N}$,
demanding that the sum of errors relative to some ``model
source'' is minimized at fixed~$N$ and $r_0 = 0$,
\beq
\sum_{j=1}^N \left| {\Delta S_j \over S^{\rm mod}_j } \right|
= {\rm min} \, ,
\label{Dmin}
\eeq
where the $\lbrace \Delta S_j \rbrace_{j =1 , \ldots , N}$ are the
square roots of the $\{\Delta^2S_j\}_{j=1,\ldots,N}$ in
Eq.~(\ref{DSl}).  We~find a~rather weak
sensitivity of the results to fine details of $S^{\rm mod}$
in~(\ref{Dmin}), so we just use a~simple exponential form
$S^{\rm mod} \propto
\exp{(-r/R_0')}$, $S^{\rm mod}_j = S^{\rm mod}((r_{j-1}
+ r_j)/2)$, with~$R_0'$ of the order of few~fm.
The~exponential
form is consistent with a~possible tail in the source due to
prolonged decays.

Features of the squared wavefunction in~(\ref{KPhi})
and the binning in q appear to have the greatest affect
on determining the best set \rj.  Nevertheless, it is
important to use {\em relative} errors, with some
sensible~$S^{\rm mod}$, in~(\ref{Dmin}).
If~absolute errors are taken, then the $r^2$ weight
from angle-averaging in~(\ref{Kij}) favors large $r$'s.
The net result is that we learn that the source is close to
zero at large $r$ to a very high accuracy; we do not need
imaging to tell us this.
Our practical observation is that the~sum of
relative errors in~(\ref{Dmin}),
rather than the sum of squares, is preferred for minimization;
the~sum of squares pushes \rj  inwards, leaving little
resolution at high~$r$.

\begin{figure}
\vspace*{.2in}
\begin{center}
\epsfxsize=5.5in 
\epsffile{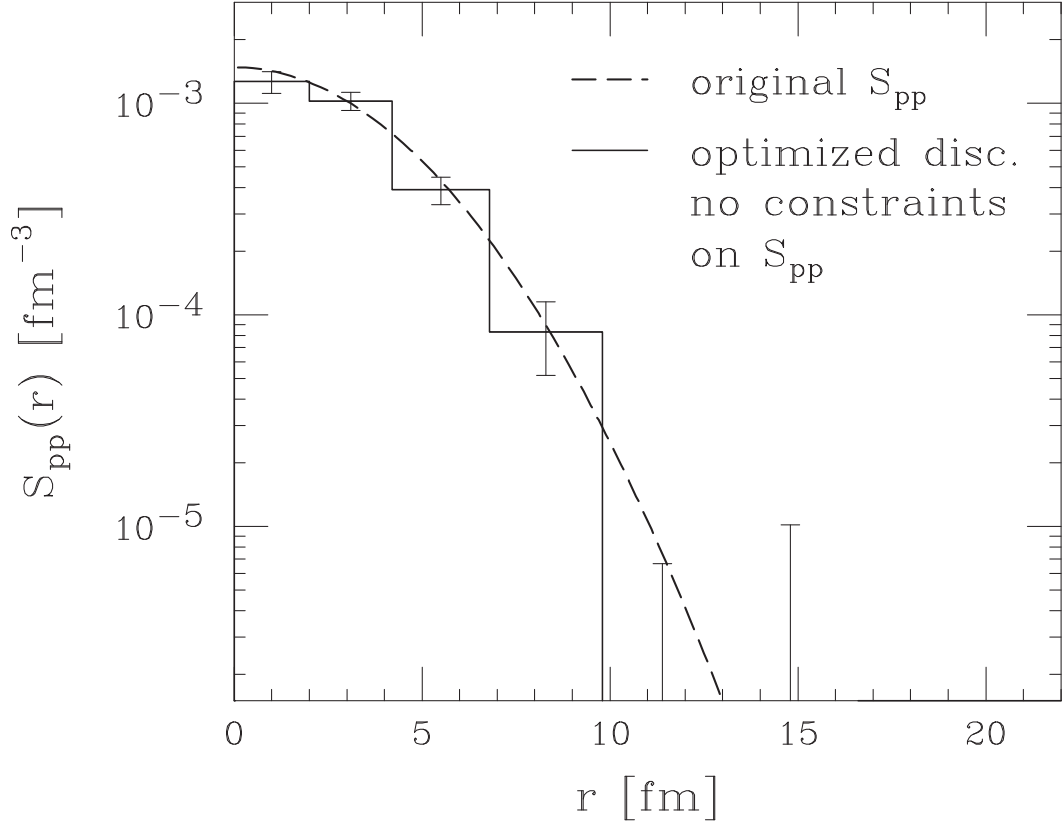}
\vspace*{.4in}
\end{center}
\caption{
This plot is the same as Figs.~\ref{nocon2} and~\ref{con2},
except that the source is not constrained and
is restored with the optimized discretization method.
}
\label{nocono}
\end{figure}

\begin{figure}
\vspace*{.2in}
\begin{center}
\epsfxsize=5.5in 
\epsffile{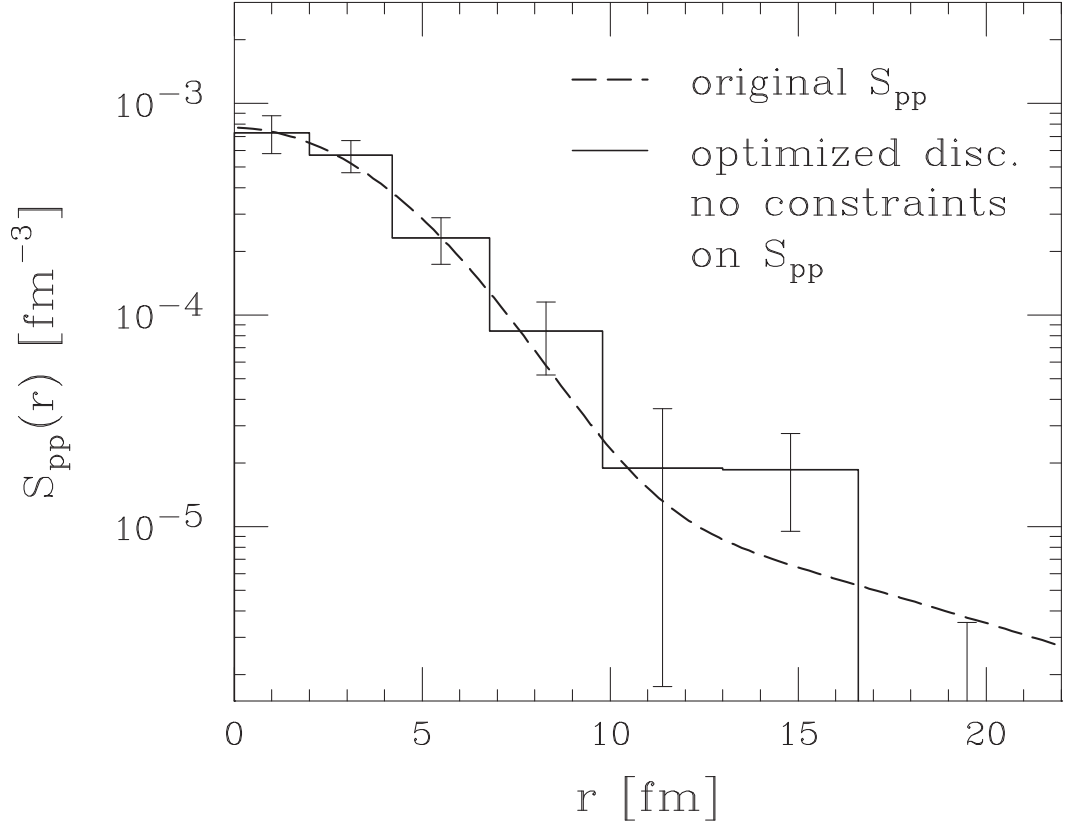}
\vspace*{.4in}
\end{center}
\caption{The solid histogram is the relative pp source
restored using the optimized discretization method.  The correlation
function used is indicated by the filled circles in
Fig.~\ref{simcor}.
The original source function is shown with the dashed line.
}
\label{noconot}
\end{figure}

Once an optimal set of \rj is found, Eq.~(\ref{S=}) can be used to
determine the source.  The results of applying  our procedure to
the simulated pp correlation function of the preceeding
section, are shown in~Fig.~\ref{nocono}, for $N=7$.
The~optimal intervals
for discretization typically increase in size with~$r$.
For example,
the first interval in Fig.~\ref{nocono} is 2~fm wide and the
sixth is 3.6~fm wide.
The figure clearly shows that we can satisfactorily restore
the source
without imposing any constraints.  Figure~\ref{noconot} shows
the results from a~similar restoration of the source with an
exponential tail:\\[-4ex]
\beq
S(r) = {1 \over 2} \, {1 \over (2 \pi R_0^2)^{3/2}}
\exp{\left(- \, {r^2 \over 2 R_0^2} \right)} + {1 \over 2} \,
{15 \over 4 \pi^5 R_1^4} \, {r \over \exp{(r/R_1)} - 1}
\, , \\[-4ex]
\label{SGo}
\eeq
where~$R_0 =3.5$~fm and $R_1 = 6$~fm.  We show the~correlation function
corresponding to the restored source in Fig.~(\ref{simcor}), both
with and without errors of the rms magnitude of~0.015.  Since~the
same~$N$ and the same $\lbrace q_i,\sigma_i
\rbrace_{i=1,\ldots,M}$ are used in the inversion,
we find the~same optimal \rj used in Fig.~\ref{nocono}.
The~restored source gives evidence for
the tail in the source, despite of the fact that the magnitude of the
tail is lower by~2 orders of magnitude compared to the
maximum at $r=0$.  Comparing figures~\ref{nocono}
and~\ref{noconot},
we see that our method can discriminate between the two source shapes.
If we impose additional constraints to the optimized discretization
method, the agreement between the restored and original source
functions further improves.

\section{Sources from pp Data}
\label{adata}

Now we apply our new optimized discretization method to
analyse the pp correlation data~\cite{gon90,gon91},
from the $^{14}$N +
$^{27}$Al reaction at 75~MeV/nucleon, imaged in
the naive
fashion in~Ref.~\cite{bro97}.
Since our new method does not require
positive  definiteness for stabilization, we are able to
lift this constraint and verify whether the data favor positive
definite sources.  Further, we do need to normalize the sources
to~1 within the imaged region.  We also compare the pp sources
from data to those from the transport model~\cite{dan95}, over
a~large range of relative separations and magnitude of the
sources.
Past experiences in comparing semiclassical transport
models to single-particle and correlation data
have been mixed~\cite{gon90,gon91,gon93,han95,gaf95}
for this particular reaction and others in this energy range.

The~low relative-momentum pp-correlations
have been determined in Ref.~\cite{gon90,gon91} for pairs
emitted
around
$\theta_{\rm lab} = 25^\circ$
from the $^{14}$N +
$^{27}$Al reaction at 75~MeV/nucleon,
in three intervals of the
total momentum:
270--390, 450--780, and 840--1230~MeV/c.  The~highest lab
momenta interval corresponds to the highest proton momenta
in the participant~cm for this reaction.  These momenta are
higher than the average for participant protons and directed
rather forward. Transport calculations~\cite{bro97,dan95}
show that the highest momenta bin is mostly populated by
pairs from the semicentral to peripheral collisions.
The~intermediate momenta interval
corresponds to the magnitude of typical momenta
of participant nucleons in the forward NN cm hemisphere.
The transport calculations show that these pairs stem mainly
from the semicentral collisions. The~lowest
lab momenta interval has both participant and target spectator
contributions.  In the latter case, the transport calculations
show that pairs are mostly from the semicentral to central
collisions.  According to the transport model, the~average
emission times for protons, from the first contact of the
nuclei, in the three momenta intervals, from the highest to
lowest, are $\sim 35$, $\sim 80$, and $\sim
110$~fm/c, respectively.

The results of analyzing data using the optimized
discretization method are presented in Table~\ref{Snco}.
While we quote only the values
obtained using wavefunctions for the REID93
potential, the~values for the NIJM2~\cite{sto94} differ
only by a~fraction of $1/1000^{\rm th}$.  Both the results with
and without the positivity and normalization
constraints are listed in the table.
The~results are similar, given their errors.
The~results obtained
without constraints are generally consistent with positive
definite source functions.  Only in the highest
total-momentum interval there might be some indication of
a~negative minimum in the source function.
One can also see in Table~\ref{Snco} that the~magnitude of $S$ in the
low-$r$ region increases by a~factor of~2 each time one
switches from a~lower to a~higher total-momentum interval.

\begin{figure}
\vspace*{.2in}
\begin{center}
\epsfxsize=6.5in 
\epsffile{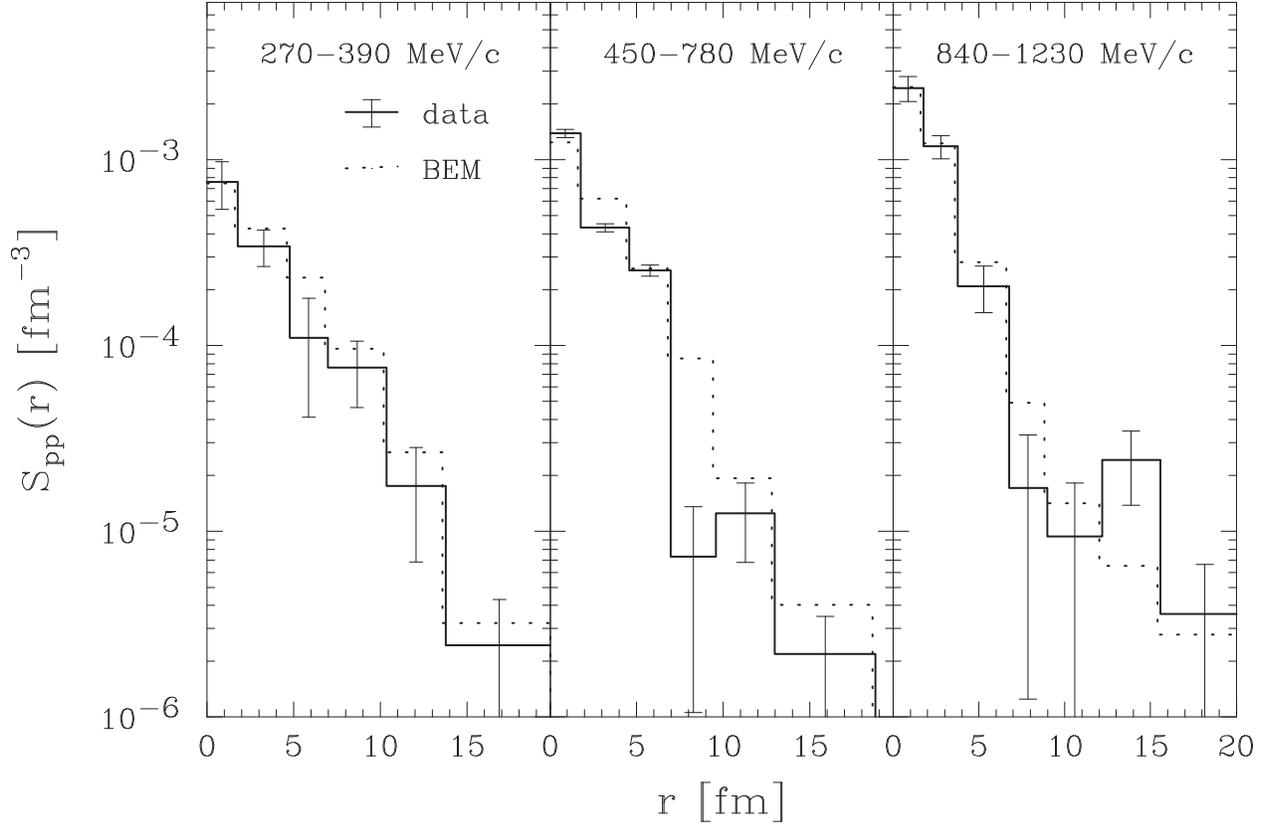}
\vspace*{.4in}
\end{center}
\caption{
Relative source
for protons emitted from the $^{14}$N + $^{27}$Al
reaction at~75~MeV/nucleon, in the vicinity of $\theta_{\rm
lab} = 25^\circ$, in the three total momentum intervals
of 270--390~MeV/c (left panel), 450--780~MeV/c
(center panel), and 840--1230~MeV/c (right panel).
Solid lines are the source values extracted from the
data~\protect\cite{gon91} and the dotted lines are the
source values obtained in the Boltzmann-equation calculation.
}
\label{pps}
\end{figure}

In~Fig.~\ref{pps}, we compare the
constrained results from the data to the distributions of
relative separation of last collision points for protons
with similar  momentum from the transport
model~\cite{bro97,dan95}.  Clearly, the~semiclassical model
can only yield positive-definite source functions.
Again, we can see the focussing of the experimental
distribution at low $r$ as the pair momentum increases.
The~large-$r$ tails in the
distribution at different momenta cannot be accommodated
with the Gaussian parametrizations used to
describe the low-$r$ behavior of the
sources~\cite{gon90,gon91}.

Generally (Fig.~\ref{pps}),
the~Boltzmann-equation model (BEM) yields relative emission
point
distributions that are similar to the imaged data, including
the dependence on total pair momentum.
In~fact,  the~maximae around
$q =20$~Mev/c (such as in Fig.~\ref{simcor} of the present
paper) are nearly the
same height as the data (see Fig.~2 in~\cite{gon93}).
Such findings are somewhat surprising
for the low and intermediate total momentum intervals.
While BEM adequately describes high-momentum wide-angle
single-particle spectra of protons, which correspond to
the highest total-momentum interval (see Fig.~1 in~\cite{gon93}),
the~model overestimates the single-particle proton spectra by as much
as 1.5--5 in the two lower momentum intervals~\cite{gon93}.
\footnote{For other comparisons of
the transport theory to single-particle data from the same or
similar reactions see~\cite{gon90,dan91,han95};
overall proton multiplicities are typically overestimated by
a~factor of~2, possibly due to excessive stopping within
semiclassical transport in this energy range.}  Looking~closer
at Fig.~\ref{pps}, we find that the distributions
from data are somewhat sharper at low-$r$ in the two lower
total-momentum
intervals than the the distributions from the model.
In the next section, we reveal a~serious
discrepancy when we go beyond a~point by point examination.

\section{Generalized Chaoticity Parameter}

An important quantity characterizing the imaged
distribution can be the integral of the source over
a region where the distribution is significant.
We introduce the parameter
\beq
\lambda (r_N) = \int_{r < r_N} d{\bf r} \, S({\bf r}) \, .
\label{lambda}
\eeq
that generalizes the
the chaoticity~$\lambda$ used to parametrize high-energy
$\pi \pi$ correlations.  The~term 'chaoticity' stems from early
models involving classical currents used to describe $\pi$
production.
The~standard chaoticity parameter
is defined by fitting the $\pi \pi$ correlation function to
a~Gaussian:
\beq
C(q) - 1 = {2 \pi \over q} \int_0^\infty dr \, r
\sin{\left(2qr \right)} S(r) \simeq \lambda \exp{\left(- \,
{2 q^2 R_0^2 } \right)} \, .
\label{Cql}
\eeq
When one uses this parametrization of the correlation function,
one assumes the following parametrization of the source at
low~$r$:
\beq
S(r) \approx { \lambda \over (2 \pi R_0^2)^{3/2}} \exp{\left(- \,
{r^2 \over 2 R_0^2} \right)}.
\label{SGl}
\eeq
Our chaoticity parameter generalizes the one in (\ref{Cql}) and
(\ref{SGl}) in two ways.
First, because our chaoticity parameter is defined in terms of
the imaged
source function, rather than a~Gaussian fit to the correlation
function, it can
apply to any particle pair, not just pion pairs.  Second it is
dependent on the cut--off, $r_N$.
The importance of our definition of the generalized chaoticity
parameter lies in the fact that some particles in the reaction,
such as pions or protons, can stem from long-lived resonances
and be emitted far from any other particles.
Thus, they contribute primarily to~$S$ at large~$r$, possibly
outside the region that may be imaged. 
If the kernel~$K$ is
close to zero or averages out to zero for moderate to high~$q$
and large~$r$, then the tail in~$S$ does not contribute to the
deviations of $C$ from~1 in Eq.~(\ref{K}) or~(\ref{Cql}).
The~imaged region is limited to~$r_N$ and may result in~$\lambda < 1$.
Access to $C$ at lower~$q$ may give higher~$r_N$ and $\lambda$
closer to~1.  However, given experimental limitations on 
measuring~$C$ at the lowest
relative~$q$, we only really stand
a~chance to study large~$r$ with charged particles
(and the~higher the charge the better).  
The charged particles with low relative~$q$ should be emitted far
apart, otherwise they acquire relative kinetic
energy due to mutual repulsion.
For the analysed
pp data~\cite{gon90}, the low-$q$ region is either not
available or is associated with large systematic errors.
Imaging can only extend up to $r_N \sim 20$~fm.  Thus,
one can expect~$\lambda (r_N) < 1$.   In many $\pi \pi$
measurements, e.g.~\cite{bar97}, the~resolution
allows one to image~\cite{bro97} regions of comparable sizes,
typically 10--20~fm.
In~contrast to the $pp$ and $\pi \pi$ data, the~data on
IMF (such as~\cite{ham96}) often extend to low values of
relative velocity because of the use of fragments with large
charges.  This permits imaging up to relative separations
as large as~50~fm.  For a~discussion of this, see the next section.

In Table~\ref{lambdapp}, we have tabulated $\lambda(r_N)$ in
each momentum
interval for the constrained and unconstrained sources of
Section~\ref{stability} as
well as sources from the BEM. Comparing the results of the BEM to the
constrained source, we only find agreement in the highest momentum interval.
It follows that, when compared to the model,  significant
portions of the source are {\em missing} from the
imaged regions.  This  discrepancy is especially
pronounced in the intermediate-momentum region.
Nevertheless, it is comforting that the
transport model describes the features
of the relative source for the high momentum protons
since it properly describes~\cite{gon93} the
high-momentum single-particle spectra.
Now, in~BEM no IMFs are
produced and the~IMFs may decay over an~extended time,
contributing to large separations in the relative emission function,
as they move away from the reaction region.
Of course these decays produce some final IMFs,
contributing to the relative IMF~sources at distances similar
to those for the pp~sources.  It~may be interesting to see
whether a~significant portion of the relative
IMF sources can extend beyond $\sim 20$~fm, as is apparent for the pp sources.
We check this in the next section of the paper.

The disagreements between the data and
calculations in both the values of $\lambda(r_N)$ and the
single-particle spectra~\cite{gon93},
for the lower momenta, reveal
unphysical features of low-momentum proton emission
in the transport model.  The coarse agreement between the
measured correlation function and the function
calculated using the model in Ref.~\cite{gon93}
in the lower total-momentum intervals, mentioned in the last
section, is coincidental.
Since the images show $\lambda(r_N)<1$, some of the strength
of~$S$ is shifted out to large $r$ lowering the height of~$C$.
For BEM, the height of~$C$ is lowered because~$S$ is softer
than $S$ from the data.
Thus, the BEM correlation function can match the
height of the measured
correlation peak, while not matching the overall shape of the
correlation function. For other systems in the
general energy range, disagreements were found even in the
height of~$C$~\cite{gon91,han95,gaf95}.

\begin{figure}
\vspace*{.2in}
\begin{center}
\epsfxsize=5.5in 
\epsffile{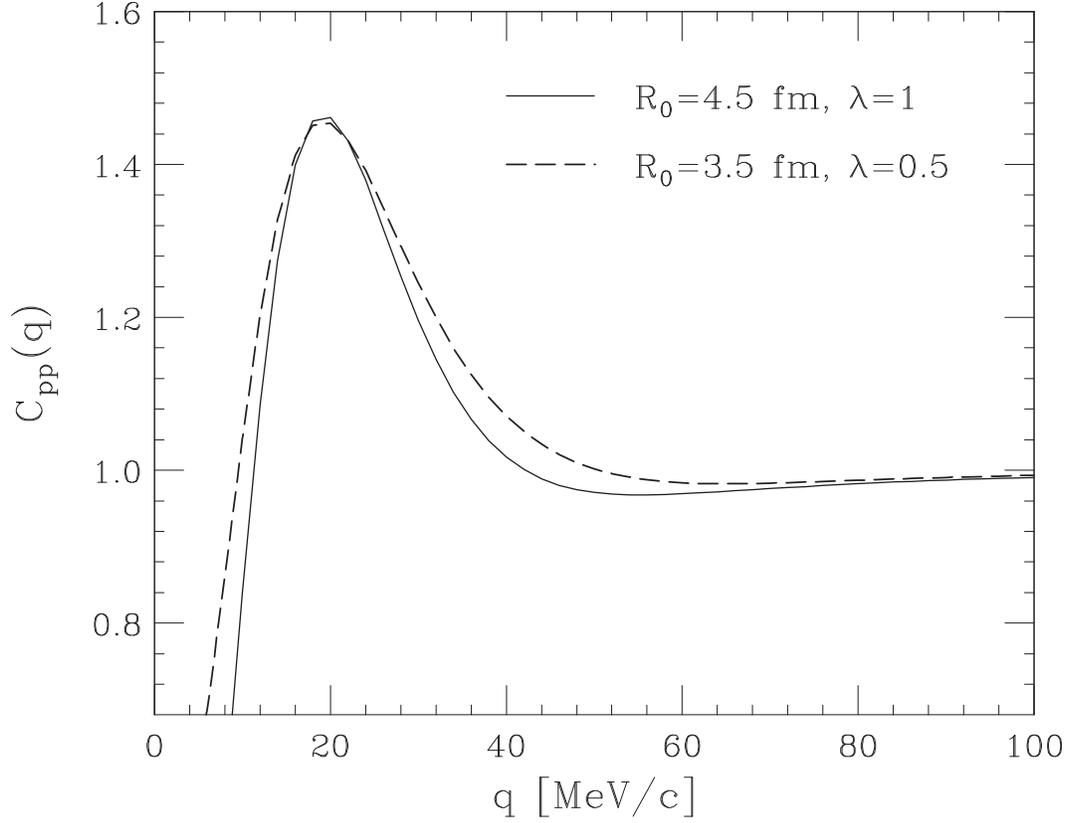}
\vspace*{.4in}
\end{center}
\caption{The solid line is the two-proton correlation function
for $R_0 = 4.5$~fm and $\lambda = 1$ while the dashed line is
for $R_0 = 3.5$~fm and $\lambda = 0.5$.
The source is the Gaussian in Eq.~(\protect\ref{SGl}).
}
\label{2cor}
\end{figure}

These conclusions raise the question of the general
sensibility of attempting to fit the magnitude of pp or nn
correlation function at the maximum~\cite{boa90,gon90,gon91}
by adjusting
the radius $R_0$ in the~Gaussian parametrization of the
source function (see Eq.~(\ref{SGl})) with~$\lambda = 1$.
When fitting the $\pi
\pi$ correlation functions at intermediate relative momenta,
{\em both} the strength and extent of the source function
at low~$r$ are varied: $\lambda$ is read off from the
magnitude of the correlation function at low~$q$ and $R_0$ is
read off from the width of the correlation function in~$q$.
However, because of the resonant nature of the low-momentum NN
interaction, the~height of the NN
correlation function is determined by the magnitude of the
source within the resonance peak of the wavefunction.
That magnitude is both affected by the overall low-$r$ strength
of the source {\em and} the low-$r$ source falloff.  This is
illustrated in Fig.~\ref{2cor} which shows pp correlation
functions for the source~(\ref{SGl}):  the~same maximum height
can be obtained using $R_0 = 4.5$~fm and $\lambda = 1$
as using $R_0 = 3.5$~fm and $\lambda=0.5$.

Given that the resonance peak in the
$^1S_0$ wavefunction is quite narrow (it has an~outer radius of
$\sim 2.5$~fm) and the source falloff cuts off large-$r$
contributions to the integration in~(\ref{K}),
the low-$r$ limit of $S$ is proportional to the
$C-1$ at the maximum (to a $\pm 20\%$ level)
for virtually all low-$r$ falloffs that may be encountered in
practice, $R_0=2.5-6.0$~fm:
\beq
C_{\bf P} (20 \, {\rm MeV/c}) \approx  1 +  540\,{\rm fm^3} \,
S_{\bf P} (r \rightarrow 0).
\label{cps}
\eeq
For the~two sources we used in Fig.~\ref{2cor}, we get
about the same value of $S(r \rightarrow 0)$ and therefore
about the same maximum height in~$C$.  With the same maximum height,
the~source falloff is reflected in the width of the
maximum in Fig.~\ref{2cor}.
Note that the value of $S$ at short relative distances
determines
the average freeze-out phase-space density and the entropy in
the reactions~\cite{ber94,bro97}; the~simple relation for
estimating the proton phase-space density takes the form, from
(\ref{cps}),
\beq
\langle f({\bf p}) \rangle = 0.0018 \, \left({ {\rm  GeV
} \over c }\right)^{3} \, ( C_{2 {\bf p}}^{\rm max} - 1) \,
{E_p \over m} \, {dN_p \over d {\bf p}} \, ,
\eeq
where $dN_p/d {\bf p}$ is the proton momentum distribution.

\section{IMF Sources}

We now turn to the~analysis of IMF sources.
We~choose the correlation data of Hamilton \etal~\cite{ham96},
from central $^{84}$Kr + $^{197}$Au reactions at
35, 55, and 70~MeV/nucleon, because these
data give us the opportunity to examine the variation of sources with
beam energy.

Pairs were collected in the angular range of $25^\circ < \theta_{\rm
lab} < 50^\circ$ in order to limit
contributions to the  correlation functions in~\cite{ham96}
from targetlike residues.  Hamilton \etal   determine the
functions in terms of the reduced velocity
\beq
v_{\rm red} = {v \over (Z_1 + Z_2)^{1/2}} \, ,
\eeq
under the assumptions that
the Coulomb correlation dominated the fragment correlation and
that the fragments were approximately symmetric,
$Z/A \approx 1/2$.   The~distance
of closest approach in~(\ref{KCou}) for symmetric fragments
is
\beq
r_c = {2 \, Z_1 \, Z_2 \, (A_1 + A_2) \, e^2 \over A_1 \, A_2
\, m_N \, v^2} = {e^2 \over m_N \, v_{\rm red}^2} \, .
\label{rc}
\eeq

\begin{figure}
\vspace*{.2in}
\begin{center}
\epsfxsize=5.5in 
\epsffile{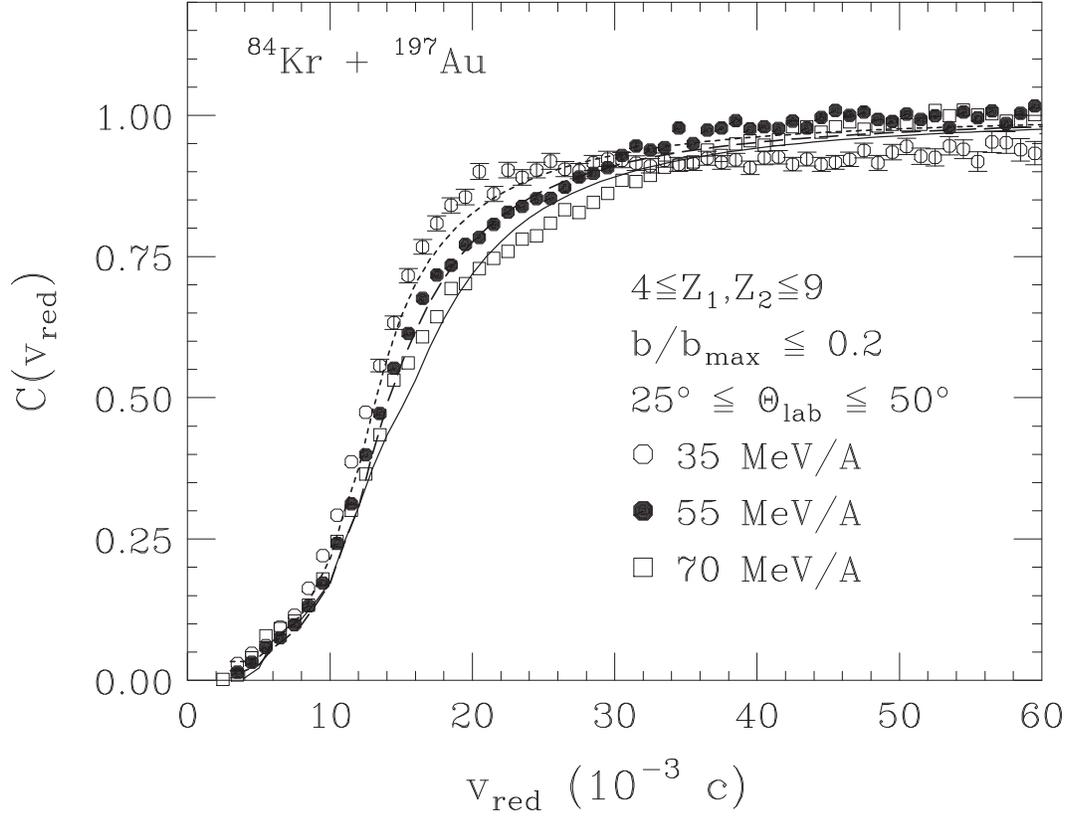}
\vspace*{.4in}
\end{center}
\caption{
Fragment-fragment velocity correlation function in central
$^{84}$Kr + $^{197}$Au reactions.  The symbols show the data of
Ref.~\protect\cite{ham96} and the lines show the
imaged source function.  The 35~MeV/nucleon data is represented by
the open circles and dotted line, the 55~MeV/nucleon data by
solid circles and dashed line, and 70~MeV/nucleon by
open squares and solid line.
}
\label{corIMF}
\end{figure}

The correlation functions at the three beam energies are shown
in Fig.~\ref{corIMF}.   We have
reduced the normalization for the correlation function at
35~MeV/nucleon by~5\%, compared to~\cite{ham96}, in order to better
satisfy the condition that $C \rightarrow
1$ at large $v_{\rm red}$, outside of the region shown in
Fig.~\ref{corIMF}.
In~examining Eqs.~(\ref{K}),
(\ref{KCou}),
(\ref{rc}), and Fig.~\ref{corIMF}, certain issues become
apparent.  Contributions to the~Coulomb
correlation
function~$C$ for a~given~$v_{\rm red}$ stem only from the
region of the source with
$r > r_c$.  As~$v_{\rm red}$ increases from~0, the~distance of
closest approach~$r_c$ decreases, with more and more inner
regions of the source~$S$ contributing to~$C$.  The~low-$v_{\rm
red}$ correlation functions at the three beam energies in
Fig.~\ref{corIMF} are quite similar, suggesting that the tails
of the source functions are similar.  Differences occur at
higher $v_{\rm red}$, indicating differences in the inner
source regions.

To image of the IMF sources, we optimize \rj as in the pp
case, but we add the constraint $r_1 \ge r_1^{\rm min}$.  We do this
because the Coulomb interaction in (\ref{K}) does not dominate
when the measured fragments are in close contact.
The Coulomb correlation alone cannot
be relied upon to get information on the most inner
portion
of the source.  The~typical touching distance for the fragments
measured in~\cite{ham96} is $r_t \sim 5$~fm; we chose a~value
$r_1^{\rm min} = 7.0$~fm which ensures that there is more
volume in the lowest bin outside $r_t$, than inside~$r_t$.

\begin{figure}
\vspace*{.2in}
\begin{center}
\epsfxsize=5.5in 
\epsffile{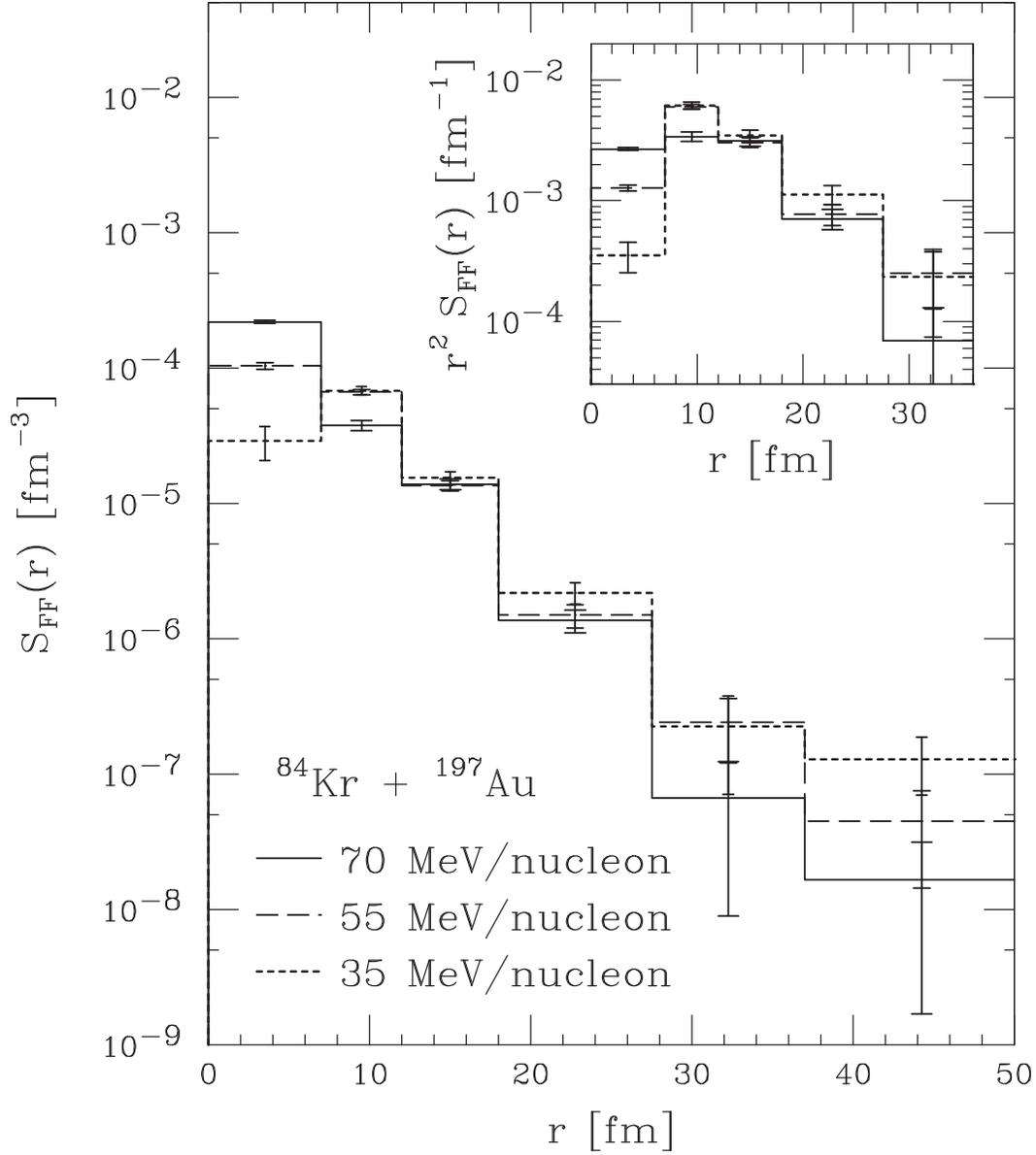}
\vspace*{.4in}
\end{center}
\caption{
Relative source for IMFs emitted from
central
$^{84}$Kr + $^{197}$Au reactions from the data of
Ref.~\protect\cite{ham96} at 35 (dotted line), 55~(dashed line), and
70~MeV/nucleon (solid line).  The insert shows the source
multiplied by~$r^2$.  In both plots, the full image extends out to $90$~fm.
}
\label{sorf}
\end{figure}

The results from the imaging are shown in Fig.~\ref{sorf}.
Given the errors in the figure, the~tails
of the sources at the three energies are not very different.
However, we observe significant variation with energy at
short relative distances ($r < 12$~fm) with the source undergoing
a~larger change between 55 and
75~MeV/nucleon, than between 35 and 55~MeV/nucleon.

In Table \ref{lambdaIMF} we tabulate the generalized chaoticity
parameter for
the IMF sources with the cutoff of 20~fm and over the whole
restored image.  The~values of $\lambda$ for the whole image
(cutoff of 90~fm) are all consistent with~1, within errors.
The~values of $\lambda$(20~fm) 20--30\% below 1 and
Fig.~\ref{sorf} indicate that a~large part of IMF emission
occurs at distances that are not imaged with protons.
Nevertheless, the~values of $\lambda (\sim$20~fm)
the~low- and high-momentum pp sources in Section
\ref{adata} are roughly equal to  the IMF $\lambda$(20~fm) but
the intermediate-momentum pp $\lambda$(20~fm) is lower.
It~should be mentioned that no complete quantitative
agreement should be expected, even if the data were from the
same reaction and pertained to the same particle-velocity
range.  This is because more protons than IMFs can stem from secondary
decays.  Besides,
the~velocity gained by a~proton in a~typical decay is large
compared to the relevant relative velocities in pp
correlations, but the
velocity gained by an~IMF can be quite small on the scale of
velocities relevant for the IMF Coulomb correlations.
Thus, the~IMF correlations may reflect
the primary parent sources, which are concentrated around the origin,
rather than the final sources.

The tails of the IMF sources extend so far that they
must be associated with the time extension of
emission.  Even so, it~is interesting to ask how far into the
center of the source must we go to see the effects due
to the spatial extent of the primary source.
In~\cite{ham96}, the combinations of single-particle source radii and lifetimes
that gave acceptable descriptions of their data
have radii varying between 5 and 12~fm.
In general, the~combination of spatial
extent and lifetime effects should give rise to a~bone-like
shape of the relative source, with the
source elongation being due to the emission lifetime.
With this, one
could try to separate the temporal and spatial effects by restoring the full
a~three-dimensional source.  In~the
angle-averaged source, the~part dominated by
lifetime effects should fall off as an~exponential
divided by the square of the separation, $r^2$,
as a function of relative separation.
On the other hand, the~part of the relative
source dominated by spatial effects may
fall off at a~slower pace or even be constant.
In~Fig.~\ref{sorf}, we see
that the~sources change weakly with~$r$ at 35 and
55~MeV/nucleon and faster at 70~MeV/nucleon,
within the range where sources vary with energy ($r \lapproxeq 12$~fm).
For reference, in the insert to Fig.~\ref{sorf}
we show the IMF source multiplied
by~$r^2$.  We see an~edge at $r \sim 11$~fm at 35 and at
55~MeV/nucleon which disappears at 70~MeV/nucleon, 
suggesting that the IMF's~are emitted from a spatial region, with a~radius
$R \sim 11/\sqrt{2} \sim 8$~fm, that becomes more dilute 
as the energy  increases.
This observation is consistant with the expectation that 
dilute sources are more effective at
emitting IMF's since IMF yields in central symmetric
collisions or from central sources in asymmetric collisions
maximize towards 100~MeV/nucleon~\cite{lyn97}.

\section{Conclusions}

We have introduced the~new method of optimized discretization to image
sources.  This~method is suitable for
determining the relative source
from different particle pair correlations in heavy-ion collisions.
Recognizing the need for stable
imaging, in our method we adjust the resolution to
minimize the relative errors of the source.  The~fact that
we can actually estimate errors in this method
gives our method a~significant advantage
over the maximum entropy method.  We tested our
method by restoring assumed {\em compact} pp sources and found
that the~quality of the restored source is comparable to the restored
source we obtain by imposing
the constraints of positivity and normalization.
Imposing these constraints in our new method further reduces
the source errors, but imposing these constraints is no longer required.
The~new method
allows one to study the {\em long-range} source structure by adjusting
the overall size of the imaged region and
the resolution at large distances.

The robustness of our new imaging method gave us the capability
to examine the positive definiteness of proton Wigner sources in
the $^{14}$N + $^{27}$Al reaction at
75~MeV/nucleon~\protect\cite{gon90,gon91}.  In~principle,
unraveling the quantal negative values of a~Wigner
function is not far fetched.  In fact, negative values of Wigner functions
have been observed in interfering atomic beams~\cite{kur97}.
Admittedly, if we had discovered such values in the heavy-ion reactions,
we would first have to check for a~possible breakdown of the assumptions
leading to~(\ref{K}) and on systematic errors in data,
before concluding on a~success.
The~extensive averaging in
the reactions\footnote{The averaging is over the impact
parameter, the~central position
for the source and emission times (${\bf R}$, and $t_1$ and
$t_2$, respectively, in Eq.~(2) in~\cite{bro97}), and over the
total pair momentum~${\bf P}$.} makes it
unlikely that a  genuinely quantal oscillation in the source
function would  survive except in the very tail of the function.

We found that our imaged
sources change significantly with the total pair momentum,
becoming sharpest for the largest momenta in the cm.
Significant portions of the imaged source are missing from the
imaged region\footnote{The imaged region corresponds to relative
distances with $r< 21$~fm.} at typical
participant momenta in the cm, but not at the highest momenta.
The chaoticity parameter (the integral of the source) from
the Boltzmann-equation model~\cite{dan91,dan95} agrees with the
data
at the highest momenta, but the integral is close to one in the
participant and target-emission momenta.
Nevertheless, the model yields the correct height of the
maximae of the correlation functions~\cite{gon93}.  This is because the
right combination of  source normalization
and sharpness in the~model can yield the right value of~$S$ at short
separations, $S_{\bf P} (r \rightarrow 0)$,
and this primarily determines the height of~$C$.
Gaussian-source fits to the height of
the pp correlation function~\cite{boa90,gon90,gon91} are of a~limited value
because
considerable source strength may lie at large relative separations.

In our analysis of midrapidity IMF sources in central $^{84}$Kr + $^{197}$Au
reactions at different beam energies, we found
a~significant variation of the sources with energy at short
distances, but not at large distances.  Considerable portions
of the IMF sources extend to large distances ($r > 20$~fm)
just like the lower total-momentum pp sources.  It~would
be very interesting to image both the IMF and pp
sources in one reaction.  There is a
deficiency of our fragmentation analysis method, namely
our method lacks
three-body Coulomb effects.  When weak, these effects could be
included as a first-order perturbation.

As this work was nearing completion, it was suggested~\cite{fel97} that
we  use nested Gaussians, of
variable width and centered at $r=0$, in place of sharp-edged
intervals in our source description.
The~Gaussians would generalize nicely to three dimensions
while maintaining the features of the error optimization in our method
and we will investigate the possibility of using them in the near future.
Our strategy
of letting the errors and the kernel choose what source they
can image is novel not just for the problem of
inverting correlations but the inversion problem in general~\cite{pre92}.

\acknowledgements
The authors thank Romualdo de Souza
for providing them the IMF correlation data in a~numerical
form. They further acknowledge discussions with Urs Wiedemann
and Hans Feldmeier.
This
work was partially supported by the National Science Foundation
under Grant PHY-9605207.

\newpage

\begin{table}
\caption{Relative pp source values restored from the data of
Ref.~\protect\cite{gon90} through the optimized
discretization method, with ($S^{\rm co}$) and without ($S^{\rm
nc}$) constraints imposed for three total momentum
gates.  The centers of the discretization intervals are
$r_{k-1/2}= (r_k + r_{k-1})/2$ and the size is $\Delta r_k =
r_k - r_{k-1}$.  The~number of intervals is either $N=6$ (two
lower momentum gates) or $N=7$ (the highest gate).
}

\newpage
\begin{tabular}{ccdr@{$\pm$}lr@{$\pm$}ld}
\multicolumn{1}{c}{$P$-Range} & k &
\multicolumn{1}{c}{$r_{k-1/2}$}
& \multicolumn{2}{c}{$10^4 \times S_k^{\rm nc}$}
& \multicolumn{2}{c}{$10^4 \times S_k^{\rm co}$}
& \multicolumn{1}{c}{$\Delta r_k$}
\\
\multicolumn{1}{c}{[MeV/c]} & &
\multicolumn{1}{c}{$[{\rm fm}]$}
&
\multicolumn{2}{c}{$[\mbox{fm}^{-3}]$}
&
\multicolumn{2}{c}{$[\mbox{fm}^{-3}]$}
& \multicolumn{1}{c}{[{\rm fm}]} \\ \hline
& 1 & 0.9 & 7.6 & 2.2 & 7.6 & 2.2 & 1.8 \\
& 2 & 3.3 & 3.55 & 0.90 & 3.42 & 0.76 & 3.0 \\
& 3 & 5.9 & 1.07 & 0.91 & 1.10 & 0.69 & 2.2 \\
270--390 & 4 & 8.7 & 0.85 & 0.34 & 0.76 & 0.30 & 3.4 \\
& 5 & 12.1 & 0.30 & 0.21 & 0.18 & 0.11 & 3.4 \\
& 6 & 16.9 & --0.024 & 0.047 & 0.024 & 0.019 & 6.2 \\
&  &  & \multicolumn{4}{c}{} &  \\
& 1 & 0.9 & 13.89 & 0.70 & 13.87 & 0.69 & 1.8 \\
& 2 & 3.2 & 4.18 & 0.24 & 4.30 & 0.21 & 2.8 \\
& 3 & 5.8 & 2.69 & 0.22 & 2.54 & 0.17 & 2.4 \\
450--780 & 4 & 8.3 & --0.10 & 0.13 & 0.073 & 0.063 & 2.6 \\
& 5 & 11.3 & 0.161 & 0.075 & 0.124 & 0.057 & 3.4 \\
& 6 & 15.9 & 0.017 & 0.018 & 0.022 & 0.013 & 5.8 \\
&  &  & \multicolumn{4}{c}{} &  \\
& 1 & 0.9 & 25.0 & 4.5 & 24.3 & 3.7 & 1.8 \\
& 2 & 2.8 & 9.8 & 2.3 & 11.8 & 1.7 & 2.0 \\
& 3 & 5.3 & 2.89 & 0.98 & 2.09 & 0.59 & 3.0 \\
840--1230 & 4 & 7.9 & --1.08 & 0.88 & 0.17 & 0.16 & 2.2 \\
& 5 & 10.6 & --0.20 & 0.38 & 0.094 & 0.088 & 3.2 \\
& 6 & 13.9 & 0.41 & 0.22 & 0.24 & 0.10 & 3.4 \\
& 7 & 18.2 & 0.009 & 0.089 & 0.036 & 0.030 & 5.2 \\
\end{tabular}

\newpage

\label{Snco}
\end{table}

\begin{table}
\caption{Comparison of the integral of the relative pp source function,
$\lambda(r_N)$, for the restored and BEM sources
in three total momentum gates in the
$^{14}$N + $^{27}$Al reaction at
75~MeV/nucleon.
The restored sources use the data of Ref.~\protect\cite{gon90}.
The integrals are truncated at the distance $r_N$.}
\vspace*{.2in}

\begin{tabular}{cr@{$\pm$}lr@{$\pm$}ldd}
\multicolumn{1}{c}{$P$-Range} &
\multicolumn{5}{c}{$\lambda(r_N)$} &
\multicolumn{1}{c}{$r_N$} \\ \cline{2-6}
\multicolumn{1}{c}{[MeV/c]} &
\multicolumn{2}{c}{unconstrained} &
\multicolumn{2}{c}{constrained} &
\multicolumn{1}{c}{BEM} &
\multicolumn{1}{c}{[fm]} \\ \hline
270-390  & 0.69  & 0.22  & 0.69  & 0.15  & 0.98 & 20.0 \\
450-780  & 0.560 & 0.065 & 0.574 & 0.053 & 0.91 & 18.8 \\
840-1230 & 0.65  & 0.37  & 0.87  & 0.14  & 0.88 & 20.8 \\
\end{tabular}

\label{lambdapp}
\end{table}

\newpage
\begin{table}
\caption{Comparison of the integrals of the midrapidity IMF
source function,
$\lambda(r_N)$,
in central $^{84}$Kr
+ $^{197}$Au reactions at three beam energies,
for different truncation points, $r_N$.
The restored sources use the data of Ref.~\protect\cite{ham96}.}
\vspace*{.2in}

\begin{tabular}{cr@{$\pm$}lr@{$\pm$}l}
\multicolumn{1}{c}{Beam Energy} &
\multicolumn{2}{c}{$\lambda(90 \,{\rm fm})$} &
\multicolumn{2}{c}{$\lambda(20 \, {\rm fm})$} \\
\multicolumn{1}{c}{[MeV/nucleon]} &
\multicolumn{2}{c}{ } &
\multicolumn{2}{c}{ }  \\ \hline
35  & 0.96 & 0.07 & 0.72 & 0.04  \\
55  & 0.97 & 0.06 & 0.78 & 0.03  \\
70  & 0.99 & 0.05 & 0.79 & 0.03  \\
\end{tabular}

\label{lambdaIMF}
\end{table}

\newpage

\end{document}